\documentclass{aa} %

\relax
\usepackage{graphicx}
\usepackage{txfonts}
\usepackage{pdflscape}
\usepackage{color}

\usepackage{amsmath}
\usepackage[utf8]{inputenc}
\DeclareUnicodeCharacter{02BC}{'}
\usepackage{amstext}
\usepackage[colorlinks=true,linkcolor=black,citecolor=blue]{hyperref}%

\begin{document} 

\title{On unveiling Buried Nuclei with JWST: a technique for hunting the most obscured galaxy nuclei from local to high redshift}

   \author{I. Garc\'ia-Bernete\inst{1,2}\fnmsep\thanks{E-mail: igbernete@gmail.com}, F.\,R. Donnan\inst{2}, D. Rigopoulou\inst{2,3}, M. Pereira-Santaella\inst{4}, E. Gonz\'alez-Alfonso\inst{5}, N. Thatte\inst{2}, S. Aalto\inst{6}, S. K\"onig\inst{6}, M. Maksymowicz-Maciata\inst{7}, M.\,W.\,R Smith\inst{8}, J.-S. Huang\inst{9}, G.\,E. Magdis\inst{10}, P.\,F. Roche\inst{2}, J. Devriendt\inst{2}, and A. Slyz\inst{2}}

   \institute{$^1$ Centro de Astrobiolog\'ia (CAB), CSIC-INTA, Camino Bajo del Castillo s/n, E-28692 Villanueva de la Can\~ada, Madrid, Spain\\
   $^2$ Department of Physics, University of Oxford, Keble Road, Oxford OX1 3RH, UK \\
  $^{3}$ School of Sciences, European University Cyprus, Diogenes street, Engomi, 1516 Nicosia, Cyprus\\
  $^4$ Instituto de F\'isica Fundamental, CSIC, Calle Serrano 123, E-28006, Madrid, Spain\\
  $^5$ Universidad de Alcal\'a, Departamento de F\'isica y Matem\'aticas, Campus Universitario, E-28871, Alcal\'a de Henares, Madrid, Spain\\
  $^6$ Department of Space, Earth and Environment, Osala Space Observatory, Chalmers University of Technology, SE-439 92 Onsala, Sweden\\
  \inst{7} School of Physics, H.H. Wills Physics Laboratory, Tyndall Avenue, University of Bristol, Bristol BS8 1TL, UK\\
  \inst{8} Department of Physics, Oliver Lodge Building, University of Liverpool, Oxford Street, Liverpool, L69 7ZE, UK\\
  \inst{9} Chinese Academy of Sciences South America Center for Astronomy, National Astronomical Observatories, CAS, Beijing, 100101, Peopleʼs Republic of China\\
  \inst{10} Cosmic Dawn Center (DAWN), Copenhagen, Denmark\\}

\titlerunning{Unveiling deeply obscured nuclei with JWST}
\authorrunning{Garc\'ia-Bernete et al.}

   \date{}

  \abstract
   {We analyze JWST NIRSpec+MIRI/MRS observations of the infrared (IR) Polycyclic Aromatic Hydrocarbon (PAH) features in the central regions ($\sim$0.26\arcsec at 6\,$\mu$m; $\sim$50-440\,pc depending on the source) of local luminous IR galaxies. In this work, we examine the effect of nuclear obscuration on the PAH features of deeply obscured nuclei, predominantly found in local luminous IR galaxies, and we compare these nuclei with ``normal'' star-forming regions. We extend previous work to include shorter wavelength PAH ratios now available with the NIRSpec+MIRI/MRS spectral range. We introduce a new diagnostic diagram for selecting deeply obscured nuclei based on the 3.3 and 6.2\,$\mu$m PAH features and$/$or mid-IR continuum ratios at $\sim$3 and 5\,$\mu$m. We find that the PAH equivalent width (EW) ratio of the brightest PAH features at shorter wavelengths (at 3.3 and 6.2\,$\mu$m) is impacted by nuclear obscuration. Although the sample of luminous IR galaxies used in this analysis is relatively small, we find that sources exhibiting a high silicate absorption feature cluster tightly in a specific region of the diagram, whereas star-forming regions experiencing lower extinction levels occupy a different area in the diagram. This demonstrates the potential of this technique to identify buried nuclei. To leverage the excellent sensitivity of the MIRI imager onboard JWST, we extend our method of identifying deeply obscured nuclei at higher redshifts using a selection of MIRI filters. Specifically, the combination of various MIRI JWST filters enables the identification of buried sources beyond the local Universe and up to z$\sim$3, where other commonly used obscuration tracers such as the 9.7\,$\mu$m silicate band, are out of the spectral range of MRS. Our results pave the way for identifying distant deeply obscured nuclei with JWST.}

      
   \maketitle
\section{Introduction}
Most galaxies contain supermassive black holes (SMBHs) at their central regions (e.g. \citealt{Hopkins05}) and thus undergo an active phase in their evolution (depending on the gas supply; e.g. \citealt{Hickox14}). There is evidence that a significant fraction of cosmic SMBH growth may be taking place in heavily obscured but intrinsically luminous AGN (e.g. \citealt{Ueda14,Mateos17}). In the local Universe, a substantial fraction of ultraluminous infrared galaxies (U/LIRGs) harbour deeply obscured nuclei (A$_{\rm V}\gg$1000; e.g. \citealt{Falstad21,Bernete22b,Donnan23}). In these extremely dusty environments, the high extinction associated with the large column densities of gas and dust generally impede their detection at many wavelengths (e.g. optical/X-ray). In deeply obscured nuclei, a mixture of a compact starburst (SB) and AGN activity might power the infrared (IR) emission but the dominant power source is still under debate (e.g. \citealt{Falstad21, Pereira-Santaella21,Bernete24b}). Moreover, these buried nuclei are considered to be an important phase of galaxy evolution (e.g. \citealt{Aalto15}).

The dust surrounding the central engine absorbs a significant part of the AGN/SB radiation and then reprocesses it to emerge in the IR (e.g.  \citealt{Pier92}). Dust silicate grains, which are an important component of interstellar dust (\citealt{Mathis77}), produce features in the IR at $\sim$9.7 and 18\,$\mu$m (e.g., \citealt{Ossenkopf92}). Buried sources show strong absorption features produced by icy material (e.g. H$_2$O, CO, CO$_2$; at $\sim$3 and 6\,$\mu$m; see \citealt{Boogert15} for a review). In particular, the {\textit{dirty}} H$_2$O ice absorption features (i.e., a mix of water and other molecules such as CO and CO$_2$) are detected in the IR spectra of obscured intermediate-luminosity AGN (\citealt{Bernete24a}), buried nuclei in U/LIRGs (e.g. \citealt{Spoon07,Spoon22,Alonso-Herrero24,Perna24,Hermosa25}) and embedded massive proto-stars (e.g. \citealt{Boogert15}).

Recent works using the unprecedented combination of high angular and spectral resolution (R$\sim$1500-3500) in the 1.0-28.1 $\mu$m range, provided by the {\textit{JWST}} Near-Infrared Spectrograph (NIRSpec; \citealt{Jakobsen22, Boker22}) and the {\textit{JWST}} Mid-Infrared Instrument (MIRI; \citealt{Rieke15, Wells15, Wright15}), are transforming our understanding of deeply obscured nuclei. These capabilities have been key to investigating Polycyclic Aromatic Hydrocarbons (PAHs) and the inner dusty structures of these regions (e.g. \citealt{Alonso-Herrero24,Bernete24c,Buiten24,Donnan24,Gonzalez-Alfonso24,Pereira24,Rigopoulou24,Hermosa25}, for some of the latest publications).

\begin{figure*}[ht!]
\centering
\includegraphics[width=18.7cm]{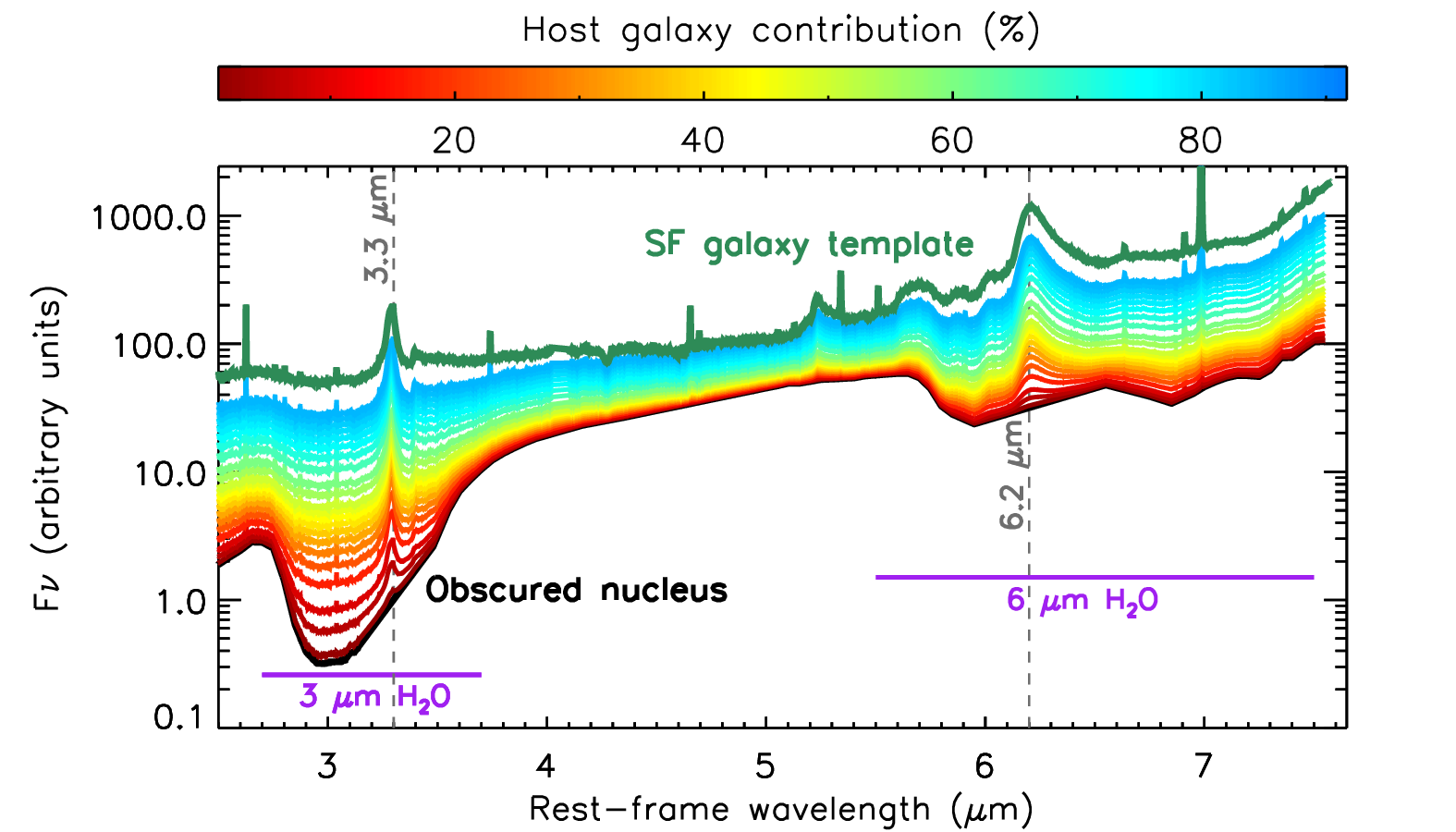}
\caption{Expected 2-to-8\,$\mu$m spectra of galaxies containing a deeply obscured nucleus with varying specific fraction relative to the host star-forming galaxy. These spectra are generated using different fractions of the host galaxy (represented by the host star-forming galaxy using the JWST spectrum of NGC\,3256-N; solid green line) superimposed on the continuum of the buried nucleus NGC\,3256-S (black solid line). Dark red to blue solid lines correspond to increasing fractions of the host galaxy contribution with respect to the nuclear source. The grey vertical dashed lines denote the location of the brightest PAH features (3.3 and 6.2\,$\mu$m) within the 2-to-8\,$\mu$m spectral region.}
\label{Synth_con}
\end{figure*}

PAHs are ubiquitous in local sources (e.g. \citealt{Tielens08} for a review), but also in high-redshift galaxies (e.g. \citealt{Spilker23}). These molecules are part of the smaller size end of the interstellar (ISM) dust distribution. PAH emission features (e.g. \citealt{Tielens08}) are very prominent in the mid-IR spectra of U/LIRGs (e.g. \citealt{Lutz98,Rigopoulou99}), and have also been detected in the IR spectra of deeply obscured sources using the relatively large aperture offered by AKARI and Spitzer (e.g. \citealt{Spoon07,Spoon22,Imanishi07,Falstad21,Bernete22b,Donnan23a}) and high angular JWST observations (e.g. \citealt{Donnan23,Donnan24,Rich23,Bernete24a,Bernete24b,Bohn24,Gonzalez-Alfonso24,Pereira24,Hermosa25}). PAHs absorb a significant fraction of UV and optical photons (e.g. \citealt{Peeters04}) and are, therefore, good tracers of the star formation activity in galaxies near and far (e.g. \citealt{Rigopoulou99,Peeters04,Shipley16}).  

In \citet{Bernete22b} a technique was introduced to identify deeply obscured nuclei based on the impact of the silicate feature on the equivalent width (EW) of the 11.3~$\mu$m PAH feature. Using Spitzer/IRS data
we demonstrated that the presence of a deeply embedded nucleus can be inferred using PAH EW(6.2~$\mu$m)/EW(11.3~$\mu$m) and PAH EW(12.7~$\mu$m)/EW(11.3~$\mu$m) ratios. Here we extend the technique using shorter wavelength PAH features (3.3 and 6.2~$\mu$m) and the 3-to-5\,$\mu$m continuum emission. We also explore the use of broad-band JWST filters to unveil embedded nuclei. The advantage of this new diagnostic is that it enables us to search for deeply obscured sources at higher redshifts up to z$\sim$3. 

Hereafter, we refer to deeply obscured nuclei as those sources with log N$_H$ (cm$^{-2}$)$>$24.0 and lacking emission lines with ionization potential ($>$40\,eV) in their nuclear IR spectrum (e.g. II\,Zw\,096-D1; see \citealt{Bernete24b}). It is also worth pointing out that the 2 to 6 $\mu$m mid-infrared (MIR) continuum of these deeply obscured nuclei rises more steeply compared to typical star-forming galaxies. Such a steep rise in the (MIR) continuum has also been noted in the Spectral Energy Distribution (SED) of the galaxies termed as Little Red Dots (LRDs; \citealt{ppg24}). LRDs have been uncovered by recent JWST surveys as a significant population of compact possibly deeply obscured galaxies at redshifts 4$<$z$<$9 whose nature is yet unclear (e.g. \citealt{Williams24, Stepney24}).

 The luminosity distance and spatial scale used in this work were calculated using a cosmology with H$_0$=70 km~s$^{-1}$~Mpc$^{-1}$, $\Omega_m$=0.3, and $\Omega_{\Lambda}$=0.7.

\begin{table*}
\caption{Properties of the targets.}
\centering
\begin{tabular}{llccccccc}
\hline
Name & z & Spatial&$\tau_{9.8\mu m}$  & EW 
(6.2/3.3) & cont (3\,$\mu$m/5\,$\mu$m) & EW 
(12.7/11.3) \\
 & &scale\\
 & &(pc arcsec$^{-1}$)\\
\hline
VV\,114-NE&0.0200& 421& 5.3& 18.3$\pm$0.9& 0.146$\pm$0.004& 0.527$\pm$0.038\\
VV\,114-S2 &0.0200& 421& 6.6& 19.1$\pm$1.0& 0.044$\pm$0.002& ...\\
IRAS\,05189-2524 &0.0441&898& 1.9& 18.1$\pm$0.9& 0.469$\pm$0.014& 2.372$\pm$0.176\\
IRAS\,07251-0248-E &0.0878& 1700& 7.0& 0.06$\pm$0.01 & 0.064$\pm$0.002& 0.024$\pm$0.002\\
IRAS\,08572+3915-NW &0.0582&1180& 6.1&... & 0.102$\pm$0.003 & 0.048$\pm$0.003\\
IRAS\,09111-1007-W &0.0541& 1110& 2.8&20.5$\pm$1.2 & 0.531$\pm$0.016& 0.612$\pm$0.043\\
NGC\,3256-N&0.0094& 193& 1.1& 84.6$\pm$4.2 & 0.459$\pm$0.011&0.868$\pm$ 0.062\\
NGC\,3256-S&0.0094& 193& 4.7& 8.8$\pm$0.6& 0.007$\pm$0.002& 0.523$\pm$0.037\\
IRAS\,13120-5453&0.0308& 650& 2.8&20.3$\pm$1.0 & 0.280$\pm$0.008& 1.056$\pm$0.074\\
IRAS\,14348-1447-NE &0.0823&1610&4.2&81.7$\pm$4.3& 0.128$\pm$0.004& 0.463$\pm$0.032\\
IRAS\,14348-1447-SW &0.0823&1610&5.2& 12.6$\pm$0.8 & 0.094$\pm$0.003& 0.162$\pm$0.01\\
Arp\,220-W & 0.0184& 394&6.3& 8.7$\pm$0.4& 0.199$\pm$0.006& 0.060$\pm$0.001\\
Arp\,220-E & 0.0184& 394&7.0& 41.9$\pm$2.2& 0.098$\pm$0.003& 0.054$\pm$0.001\\
II\,Zw\,096-D1&0.0362& 783& 2.3& 8.0$\pm$0.5& 0.153$\pm$0.004& ...\\
IRAS\,22491-1808-E &0.0778& 1500&5.8&  14.9$\pm$0.7& 0.130$\pm$0.004& 0.678$\pm$0.048\\
\hline
SF average from NGC\,7469 &0.0163& 349& 1.3&78.61$\pm$2.87& 0.677$\pm$0.091 &0.803$\pm$0.181\\
Circumnuclear region of NGC\,3256 &0.0094& 193& 2.0&76.72$\pm$4.60& 0.481$\pm$0.015 &0.690$\pm$0.048\\
\hline
\end{tabular}                                    
\tablefoot{
The redshift, and spatial scale were taken from the NASA/IPAC Extragalactic Database (NED). $\tau_{9.8\mu m}$ corresponds to the silicate strength from \citet{Donnan24}. Given that the IR emission of VV\,114-S2 is not detected at $\lambda>$10\,$\mu$m, the $\tau_{9.8\mu m}$ is derived from $\tau_{6.0\mu m}$ reported in \citet{Gonzalez-Alfonso24} by using the dust model used in \citet{Efstathiou95}. PAH EW and continuum ratios are derived in this work (see Section \ref{tool}). The EW(12.7/11.3) values are derived using the IR fitting tool described in \citet{Donnan24}.
}
\label{table_prop}
\end{table*}

\section{Targets and observations}
\label{sample}
This study aims to extend the PAH EW diagnostic diagram for identifying deeply obscured nuclei presented in \citealt{Bernete22b} to shorter wavelength PAH features, which enables the search for these sources at higher redshift with JWST. 

We use NIRSpec IFU and MIRI MRS data of local luminous IR galaxies from the Director’s Discretionary Early Release Science Program \#1328 (PI: L. Armus \& A. Evans), the Guaranteed Time Observations Program \#1267 (PI: D. Dicken; Program lead: T. Böker) and the GO Cycle 2 Large Program \#3368 (PI: L. Armus \& A. Evans). We refer to \citet{Pereira24} for further details on the sample (see also Table \ref{table_prop}). 

\section{JWST data reduction}
\label{reduction}
We retrieved near-IR to mid-IR (2.87-5.27~$\mu$m) data observed using  integral-field spectrographs MIRI MRS with a spectral resolution of R$\sim$3700--1300 (\citealt{Labiano21}) and NIRSpec with the grating-filter pairs G140H (0.97–1.89~$\mu$m), G235H (1.66–3.17~$\mu$m) and G395H (2.87–5.27~$\mu$m) with R$\sim$2700 (\citealt{Jakobsen22,Boker22}). The NIRSpec and MRS data of these sources have already been presented in previous works (e.g. \citealt{Bernete24b,Bianchin24,Bohn24,Buiten24,Donnan24,Donnan24b,Pereira24,Rigopoulou24} and references therein for some recent works).

For data reduction, we used the JWST calibration pipeline (version 1.12.4; \citealt{Bushouse23}) and the Calibration Reference Data System (CRDS) context 1253. We primarily followed the standard MRS pipeline procedure (e.g. \citealt{Labiano16} and references therein) and the same configuration of the pipeline stages described in \citet{Bernete22d} and \citet{Pereira22} to reduce the data. Some hot and cold pixels are not identified by the current pipeline, so we added some extra steps as described in \citet{Pereira24} and \citet{Bernete24a} for NIRSpec and MRS, respectively.

We extracted the central spectra by applying a 2D Gaussian model as described in \citet{Bernete24a,Bernete24b}. To do so, we employed observations of calibration point sources (MRS HD-163466 and IRAS\,05248$-$7007, Programme IDs 1050 and 1049) to measure the width and position angle of a 2D Gaussian for each spectral channel. To obtain the point source flux we used the models of the calibration PSF stars from \citet{Bohlin20}, which is equivalent to applying aperture correction factors. We refer the reader to \citet{Bernete24a,Bernete24b,Bernete24c} for further details. We extracted the circumnuclear spectrum of NGC\,3256-N by using an annulus with inner radius of 0.7\arcsec and outer radius of 1.0\arcsec. For the star-forming regions of NGC\,7469, we used the apertures from \citet{Donnan24}, excluding those located along the AGN-outflow direction (e.g. \citealt{Rigopoulou24}), as they are affected by it.

\section{The 2-to-8\,$\mu$m emission of deeply obscured galaxy nuclei}
\label{SED}
The emerging IR spectrum of galaxies containing a buried nuclei is the superposition of two main components (see e.g. Fig. 1 of \citealt{Bernete22b}): the star-forming circumnuclear regions, which are the primary sources of PAH emission; and b) the obscured nucleus which is dominated by nuclear dust emission and deep absorption features (e.g. \citealt{Marshall18,Bernete22b}). Due to the extinction law (e.g. \citealt{Ossenkopf92}), the continuum slope becomes increasingly pronounced at shorter wavelengths when extinction is high as in deeply obscured nuclei (see Fig. \ref{Synth_con}). Dusty molecular features such as silicates (at 9.7 and 18\,$\mu$m) and {\textit{dirty}} H$_2$O ice (at 3 and 6\,$\mu$m) are also generally present in buried nuclei (e.g. \citealt{Spoon07,Imanishi07,Spoon22,Bernete22b,Bernete24a}). Therefore, the specific shape of the nuclear continuum of the deeply obscured nuclei has a different impact on the EW of the various PAH features distributed across the IR waveband (main features at 3.3, 6.2, 7.7, 8.6, 11.3, 12.7 and 17~$\mu$m; see e.g. \citealt{Armus07,Desai07,Spoon07,Imanishi07,Marshall18,Hernan-Caballero20,Bernete22b}).

\begin{figure}
\centering
\par{
\includegraphics[width=10.0cm]{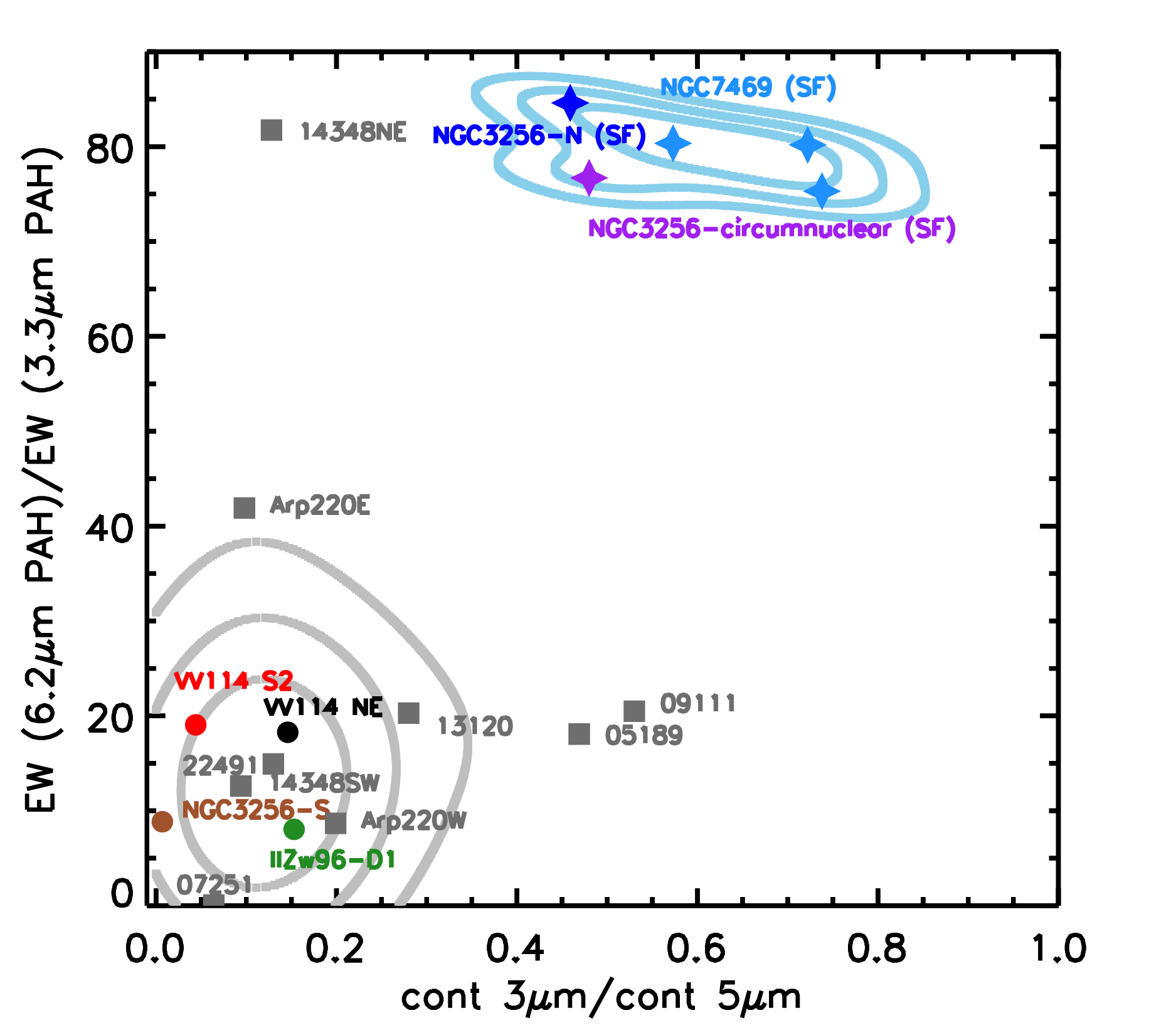}
\par}
\caption{Diagnostic diagrams for identifying deeply obscured nuclei: 6.2/3.3\,$\mu$m PAH EW ratio versus the continuum 3-to-5\,$\mu$m flux ratio. Filled circles correspond to buried nuclei in local LIRGs, and filled grey squares correspond to local ULIRGs. Dark blue and purple stars represent the central and circumnuclear regions of the star-forming galaxy NGC\,3256, respectively, while the light blue stars indicate star-forming regions in NGC\,7469. Labelled sources correspond to the the various nuclei used in this work. Grey and light blue solid lines denote the 1$\sigma$, 2$\sigma$ and 3$\sigma$ contours for deeply obscured nuclei and star-forming regions, respectively.}
\label{ratio_diagram_eqw}
\end{figure}

Previous works found that the foreground extinction may dominante the measured silicate strength of edge-on galaxies (e.g. \citealt{Goulding12,Gonzalez-Martin13}). Thus, the silicate strength alone is not necessarily a good indicator of nuclear obscuration (e.g. \citealt{Alonso-Herrero11,Gonzalez-Martin13,Bernete19,Bernete22c,Bernete24a}). However, for buried sources, \citet{Bernete22b} 
demonstrated that the nuclear 9.7\,$\mu$m silicate absorption band has a particularly pronounced effect on the 11.3\,$\mu$m PAH feature. The low flux level in the nuclear silicate absorption feature enhances the 11.3\,$\mu$m PAH band contrast (high 11.3\,$\mu$m PAH EW) compared to that of the other PAH features. Indeed, a step forward in the identification of buried nuclei was attained with the PAH EW method, but this study was limited by the limited sensitivity of Spitzer/IRS. While this method together with the unprecedented sensitivity afforded by the JWST enable the identification of buried nuclei in fainter galaxies, still the highest redshift for which this method could be applied is z$\sim$1 due to the spectral coverage (5-28\,$\mu$m) of JWST/MRS. Thus, it is 
imperative to explore alternatives for expending the mid-IR based method to higher redshifts where luminous IR galaxies are expected to be more numerous.

\begin{figure}[ht!]
\centering
\includegraphics[width=9.7cm]{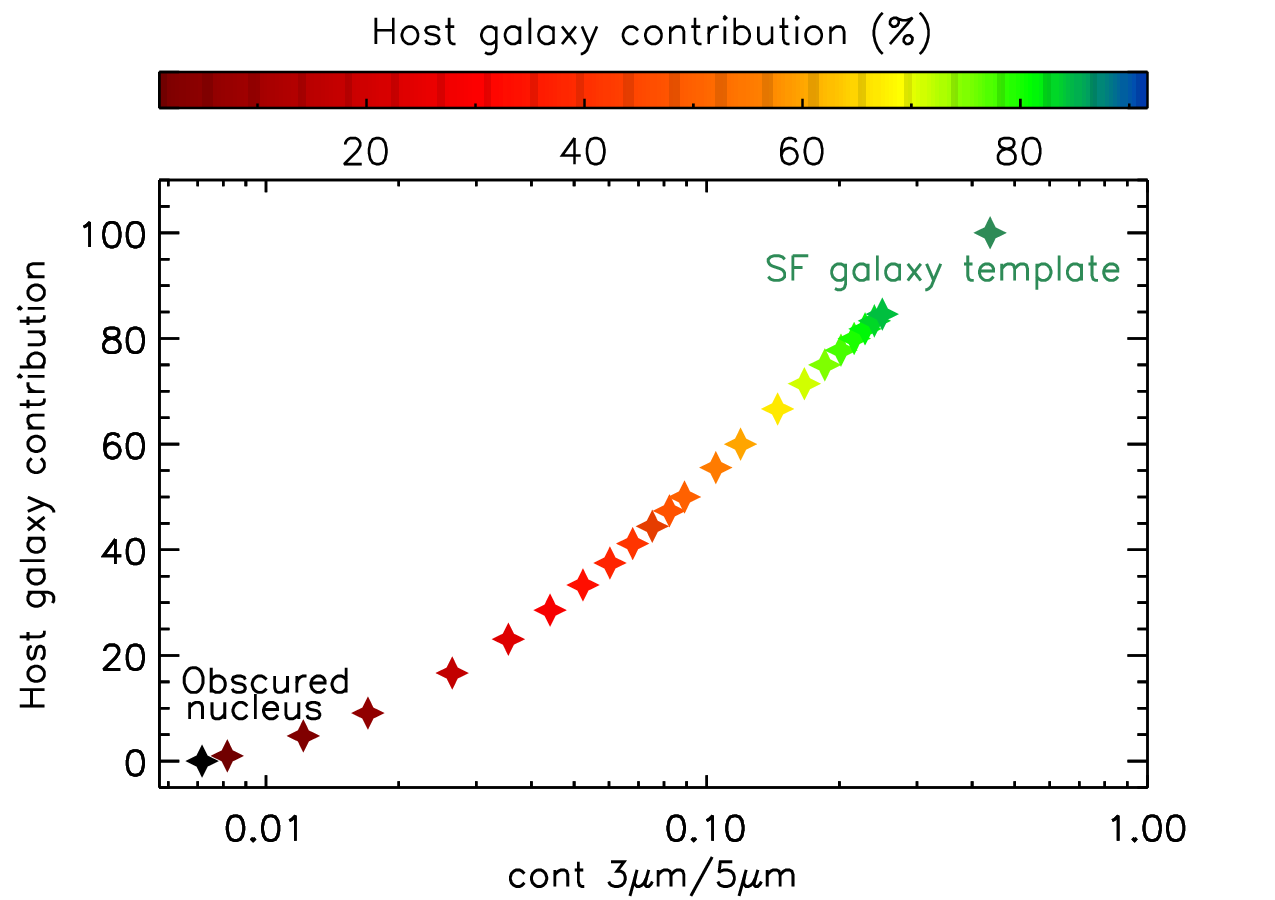}
\caption{Continuum 3-to-5,$\mu$m flux ratio (X-axis in log scale) as a function of the host galaxy contribution (Y-axis in linear scale). Colour-coded stars correspond to the host galaxy contribution of the simulated spectra shown in Fig. \ref{Synth_con}.} 
\label{contrib}
\end{figure}

For this purpose, we investigate the diagnostic power of additional spectral features at shorter wavelengths using JWST/NIRSpec$+$MRS data. In Fig. \ref{Synth_con} we generate the expected $\sim$2-to-8\,$\mu$m spectra of galaxies containing deeply obscured nuclei by using a combination of two spectra: the continuum (emission line free) spectrum of a buried nucleus whose mid-IR continuum (from NGC\,3256-S) is dominated by deep $\sim$3 and 6\,$\mu$m H$_2$O ice absorption features (black solid line), and a star-forming galaxy (from NGC\,3256-N) that represents the circumnuclear star-forming regions, that are PAH dominated (green solid line). The expected range of spectra is generated by varying the contribution of the host galaxy spectrum (from dark red to blue corresponding to increasing values of the host galaxy contribution) with respect to the nuclear source\footnote{The star-forming galaxy template is scaled by a factor to ensure that the total PAH EW remains unchanged when including a minimal contribution (1\%) from the obscured nucleus.}. This plot illustrates the variations in the $\sim$3-to-5\,$\mu$m SED slope and the effect of the deep 3\,$\mu$m H$_2$O ice absorption feature on the 3.3\,$\mu$m PAH feature. We refer the reader to \citet{Rigopoulou24} where the authors discussed the impact of the 3\,$\mu$m H$_2$O ice absorption feature on the 3.3\,$\mu$m PAH emission. Given the shape of the continuum of a buried nucleus, when one is present in a galaxy, the 3.3\,$\mu$m PAH will experience lower levels of dilution (i.e. the EW of the 3.3\,$\mu$m PAH is high) compared to other PAH features such as the 6.2~$\mu$m one. This makes this region of the spectrum an ideal wavelength range for studying deeply obscured nuclei.

\section{On the identification of deeply obscured nuclei}
\label{tool}
In this section, we investigate the relation between PAH emission and nuclear dust obscuration in our targets by plotting the 3.3/6.2~$\mu$m PAH EW ratio versus the underlying continuum ratio (Fig. \ref{ratio_diagram_eqw}). In Table \ref{table_prop} we present the PAH EW and continuum ratios of the sample; these values  were estimated using local continuum, and we used the $\sim$3.2-3.35\,$\mu$m and 6.0–6.5\,$\mu$m ranges for measuring the strengths and EWs of the 3.3\footnote{For the 3.3\,$\mu$m measurements we mask the H$_2$ 1-0 O(5) and H recombination line Pf$\delta$ lines at rest frame 3.24 at 3.29$\mu$m. We also mask gas-phase water rovibrational lines when present.} and 6.2\,$\mu$m PAH features, respectively (see \citealt{Bernete22d} for further details). We follow a similar methodology as presented in \citet{Hernan-Caballero11}.

We find that deeply obscured nuclei (filled circles) show smaller 6.2/3.3~$\mu$m PAH EW ratios than those observed in star-forming regions (filled stars). In buried nuclei, the bulk of the PAH emission, which is located in the circumnuclear star-forming regions, and the majority of the the continuum, which is coming from the nuclei, are likely to experience significantly different degrees of extinction. The EW indicates the strength of a feature compared to its underlying continuum. Thus, when a buried nucleus is present, the PAH EW ratios reflect differences in the continuum that is coming from the nuclear source (see e.g. \citealt{Bernete22b}). As the extinction curve rises sharply from $\sim$5\,$\mu$m to shorter IR wavelengths,
a significantly higher extinction value is anticipated at 3\,$\mu$m (e.g. \citealt{Ossenkopf92,Chiar06,Donnan24}). Thus, in dusty embedded nuclei, there is a very steep $\sim$3-to-5\,$\mu$m continuum slope together with deep H$_2$O absorption features. Incidentally, the $\sim$3 and 6\,$\mu$m {\textit{dirty}} H$_2$O ice absorption bands are within the wavelength range of the 3.3 and 6.2~$\mu$m PAH features. However, the nuclear 3\,$\mu$m H$_2$O ice feature band is expected to be much deeper than the 6\,$\mu$m H$_2$O (as expected from lab works; see \citealt{Rocha22} and Figs. C.1. and C.2 in \citealt{Bernete24a}), and, thus, the 6.2/3.3~$\mu$m PAH EW ratio should be smaller in buried nuclei. We also find that the 6.2/3.3~$\mu$m PAH EW values of star-forming regions are more concentrated around $\sim$75-85, suggesting that star-forming galaxies share similar intrinsic PAH flux ratios (see also \citealt{Hernan-Caballero20}). 

As demonstrated in \citealt{Bernete22b}, the PAH EW ratios filter out sources with deep absorption features due to foreground components, as these foreground absorbers equally impact both the continuum and PAH emission of the host galaxy. On the other hand, the continuum ratios are affected both by the extinction coming from the circumnuclear and nuclear region. This limitation might be present in Arp\,220-E and IRAS\,14348-1447-NE, which exhibit high PAH EW(6.2~$\mu$m)/EW(3.3~$\mu$m) ratios (similar to those of star-forming regions; see Fig. \ref{ratio_diagram_eqw}) but low 3\,$\mu$m/5\,$\mu$m continuum ratios. This suggests that foreground extinction could be playing a significant role in these sources, affecting the continuum ratio (see Appendix \ref{comparison} for further details). However, when PAH spectroscopy is not available, the ratio of the 3-to-5 $\mu$m continuum can also be used to identify deeply obscured nuclei. Because of the large dust column density, buried nuclei show a pronounced bump due to the mid-IR emission being higher relative to the low near-IR continuum. Thus we explore the ratio between the continuum at 3 and 5~$\mu$m, which appears markedly different in deeply obscured nuclei (see Fig. \ref{contrib}). Fig. \ref{contrib} shows a nearly 2 dex variation of the 3-to-5~$\mu$m continuum ratio ranging from sources dominated by deeply obscured nuclei to star-forming regions.  

We find that ratios of PAH EW(6.2~$\mu$m)/EW(3.3~$\mu$m)$\lesssim$40 and 3\,$\mu$m/5\,$\mu$m continuum ratio $\lesssim$0.35 correspond with deeply obscured nuclei in a small sample of luminous IR galaxies (Fig. \ref{ratio_diagram_eqw}). See Appendix \ref{comparison} for a comparison with the PAH EW(12.7~$\mu$m)/EW(11.3~$\mu$m) ratio method. It is worth noting that Fig. \ref{ratio_diagram_eqw} also shows that the continuum ratio alone could potentially be used to identify deeply obscured nuclei. We note, that new upcoming JWST observations of star-forming and luminous IR galaxies will be crucial for extending the sample of sources helping to establish a firm limiting criteria on the observed ratios and for precisely determining the position of star-forming regions within the diagram.

\section{Application to higher redshift sources with the JWST}
\label{filters_sect}
JWST/MIRI can observe the $\sim$3-to-5\,$\mu$m continuum range up to a redshift of $\sim$3 depending on the brightness of the source. Although the continuum emission at $\sim$3-to-5\,$\mu$m from buried nuclei may also include contributions from the host galaxy, which can even dominate in edge-on galaxies, it is of interest to explore methods to identify buried nuclei using the continuum. In particular, broad-band JWST filters can be used to identify deeply obscured nuclei candidates in distant and faint sources where spectroscopic observations might not be currently possible due to sensitivity limitations. To do so, we use the synthetic spectra\footnote{We note that we assume that the dust emission of these sources remains broadly similar at redshift 3. This assumption is consistent with recent JWST findings, which indicate that the nuclear dusty material of QSOs at $\sim$760 million years after the Big Bang is similar to that observed in local AGN (e.g. \citealt{Bosman24}).} presented in Fig. \ref{Synth_con} and combinations of the various JWST/MIRI filter transmission curves. We identify pairs of filters that allow us to use the continuum ratio described in Section \ref{tool}.

\begin{figure}[ht!]
\centering
\includegraphics[width=9.7cm]{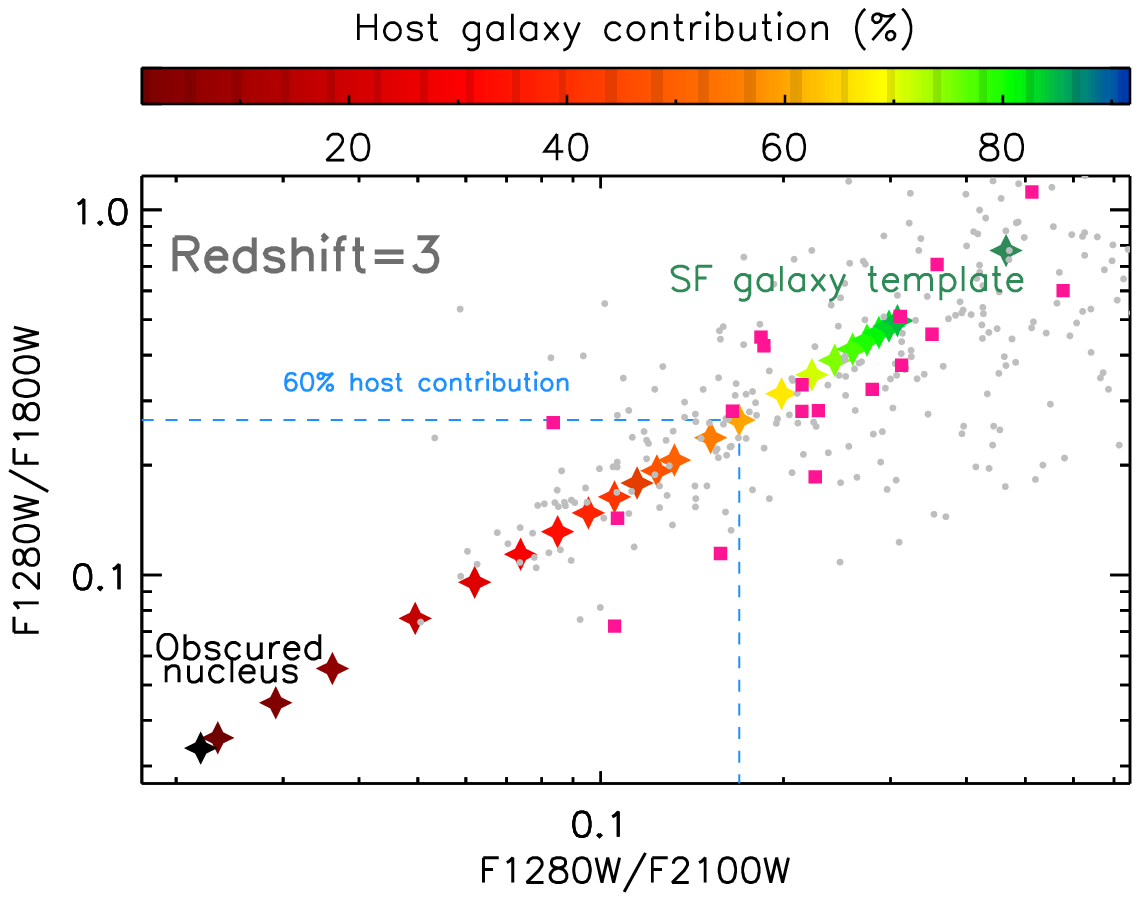}
\caption{Colour-colour diagram (using the continuum) for selecting deeply obscured nuclei at z$=3$ using JWST/MIRI filters. Colour-coded stars correspond to the host galaxy contribution of the simulated spectra shown in Fig. \ref{Synth_con}. The blue dashed lines denote the region where sources have a contribution of the buried nuclei greater than 40\%. Grey circles correspond to CEERS sources with redshift measurements. Magenta squares represent those sources at redshift $\sim$2.75-3.25. } 
\label{filters}
\end{figure}

In Fig. \ref{filters} we present a JWST colour--colour diagram based on ratios of the F1280W, F1800W and F2100W filters at z$=$3 (at rest-frame $\sim$3, 4.5 and 5\,$\mu$m, respectively). As even in heavily obscured sources a non-negligible contribution of the host galaxy is expected, we plot the expected values of the predicted spectra from Fig. \ref{Synth_con}. This plot shows that using the following criteria: F1280W/F2100W$<$0.17 and F1280W/F2100W$<$0.27, we can isolate galaxies containing a buried nuclei with more than $\sim$40\% of contribution to the observed spectrum based on Fig. \ref{contrib}. We also examine the combination of additional JWST filters that will enable us to select deeply obscured nuclei at various redshifts (see Appendix \ref{additional_filters}).

As a ``proof of concept'' for our technique, we utilise observations from the
Cosmic Evolution Early Release Science Survey (CEERS\footnote{https://ceers.github.io/}) which is part of the Director’s Discretionary Early Release Science Program 1345 (PI: Steven L. Finkelstein). The CEERS survey, which targets the Extended Groth Strip (EGS) field, includes eight MIRI pointings with a 
field-of-view (FoV) Of $\sim$2 arcmin$^2$. Two of the MIRI pointings have been observed with six MIRI filters (F770W, F1000W, F1280W, F1500W, F1800W and F2100W). Incidentally, those two MIRI pointings do not have NIRCAM observations (neither MIRI F560W). Although the surveyed area is relatively small ($\sim$ 4\,arcmin$^2$), the deep imaging available makes this an ideal target area for identifying deeply obscured nuclei candidates. To do so, we used TOPCAT (\citealt{Taylor05}) to cross-match the CEERS sources from their public catalog (\citealt{Yang23}) with a redshift catalog including both spectroscopic and photometric measurements from \citet{Kodra23}. The results are plotted in Fig. \ref{filters} (see also Appendix \ref{additional_filters}). We identify a total of 10 deeply obscured nuclei candidates within the redshift range $\sim$1.5-3 in $\sim$4 arcmin$^2$ (see Figs. \ref{filters} and \ref{filters_variousz} for different plots at z$\sim$1.5, 2, and 3). Uncertainties in photometric redshifts is an issue in this selection, and further refinement including additional photometric data (e.g. those from future NIRcam observations), will allow for more accurate redshift determination. As discussed above, strong foreground extinction may also mimic these continuum ratios, so spectroscopic follow-up of these candidates will be necessary to firmly confirm the deeply obscured nature of their nuclei. 

We note that other mid-IR selected buried sources such as hot dust-obscured galaxies (hot DOGs) are found in an even lower number density (one for every $\sim$30\,deg$^2$; \citealt{Assef15}) compared with the deeply obscured nuclei selected in this work ($\sim$2-3 candidates per arcmin$^2$). It is however, unclear what is the link (evolutionary or otherwise) between the deeply obscured sources identified through our technique and other deeply obscured populations such as hot DOGs and Little Red Dots (LRDs). Such an investigation will be presented in a forthcoming paper (\textcolor{blue}{Rigopoulou et al. in prep.}). 

In the near future, as additional JWST deep imaging observations become available, there will be large amount of archival MIRI imaging data  for studies like the one presented here. This will allow to uncover more deeply obscured nuclei at high redshifts.

\section{Summary and conclusions}
\label{conclusions}
We present a {\textit{JWST} NIRSpec+MIRI/MRS study on the relation between the PAH emission and nuclear dust obscuration in local luminous IR galaxies. We extend the work presented in \citet{Bernete22b} to include shorter wavelength PAH ratios now available with the NIRSpec+MIRI/MRS spectral range. Our main results are as follows: 

\begin{enumerate}

\item We find that the PAH EW ratio of the brightest PAH features at shorter wavelengths (i.e. 3.3 and 6.2\,$\mu$m) is related to nuclear obscuration. For deeply obscured nuclei, we observe that the slope of the intrinsic continuum and the 3\,$\mu$m water ice absorption band, which traces obscuration, have a significant impact on the EW of the 3.3\,$\mu$m PAH feature. We find that the 3.3\,$\mu$m PAH feature contrast (high EW) relative to other PAH features (e.g. 6.2\,$\mu$m) traces the nuclear obscuration. \\
 
\item We introduce a diagnostic diagram for selecting deeply obscured nuclei based on the 3.3 and 6.2\,$\mu$m PAH features and mid-IR 3-to-5\,$\mu$m continuum ratios. We find that highly obscured sources cluster tightly in a specific region of the diagram, whereas star-forming regions experiencing lower extinction levels occupy a different area. \\

\item We find that the use of mid-IR colour–colour diagrams might be effective tool to select sources dominated by deeply obscured nuclei at higher redshifts. In particular, the combination of various MIRI JWST filters will enable the identification of buried sources up to z$\sim$3.\\

\item Using the Cosmic Evolution Early Release Science Survey (CEERS) we successfully identify deeply obscured nuclei candidates within the redshift range $\sim$1.5-3.\\

\end{enumerate}

In summary, this work employs JWST/NIRSpec+MRS spectroscopy to demonstrate the potential of PAH features at 3.3 and 6.2\,$\mu$m for identifying buried sources, especially at high redshift (z$\sim$3), where other commonly used tracers of obscuration in the local universe (e.g. the 9.7\,$\mu$m silicate absorption band) might not be available due to the IR spectral range covered by current facilities. New observations of luminous IR galaxies with JWST will be crucial for enhancing the statistical significance of the results presented here, helping to establish firm criteria in the diagram.

\begin{acknowledgements}
The authors acknowledge the GOALS DD-ERS team for developing their observing program. IGB is supported by the Programa Atracci\'on de Talento Investigador ``C\'esar Nombela'' via grant 2023-T1/TEC-29030 funded by the Community of Madrid. MPS acknowledges support under grants RYC2021-033094-I and CNS2023-145506 funded by MCIN/AEI/10.13039/501100011033 and the European Union NextGenerationEU/PRTR. S. K. gratefully acknowledges funding from the European Research Council (ERC) under the European Union’s Horizon 2020 research and innovation programme (grant agreement No. 789410). 

This work is based on observations made with the NASA/ESA/CSA James Webb Space Telescope. The data were obtained from the Mikulski Archive for Space Telescopes at the Space Telescope Science Institute, which is operated  by the Association of Universities for Research in Astronomy, Inc., under  NASA contract NAS 5-03127 for JWST; and from the European JWST archive (eJWST) operated by the ESAC Science Data Centre (ESDC) of the European Space Agency. These observations are associated with the programs \#1049, \#1050, \#1267, \#1328 and \#3368. The authors are extremely grateful the {\textit{JWST}} helpdesk for their support. Finally, we thank the anonymous referee for their useful comments.

\end{acknowledgements}


\begin{appendix}
\section{Comparison with the PAH EW(12.7\,$\mu$m)/EW(11.3\,$\mu$m)  ratio}
\label{comparison}
In this Section, we compare the results from the PAH EW(6.2\,$\mu$m)/EW(3.3\,$\mu$m) ratio method presented in this work with the PAH EW(12.7\,$\mu$m)/EW(11.3\,$\mu$m)  ratio used in \citet{Bernete22b}. In general, we find a good a agreement between both methods the only exceptions are Arp\,220-E and IRAS\,14348-1447-NE. However, these two sources exhibit relatively low HCN-vib emission compared to the other nuclei in their systems (\citealt{Falstad21}), suggesting that their nature is less obscured compared to compact obscured nuclei.

\begin{figure}[ht!]
\centering
\par{
\includegraphics[width=6.7cm]{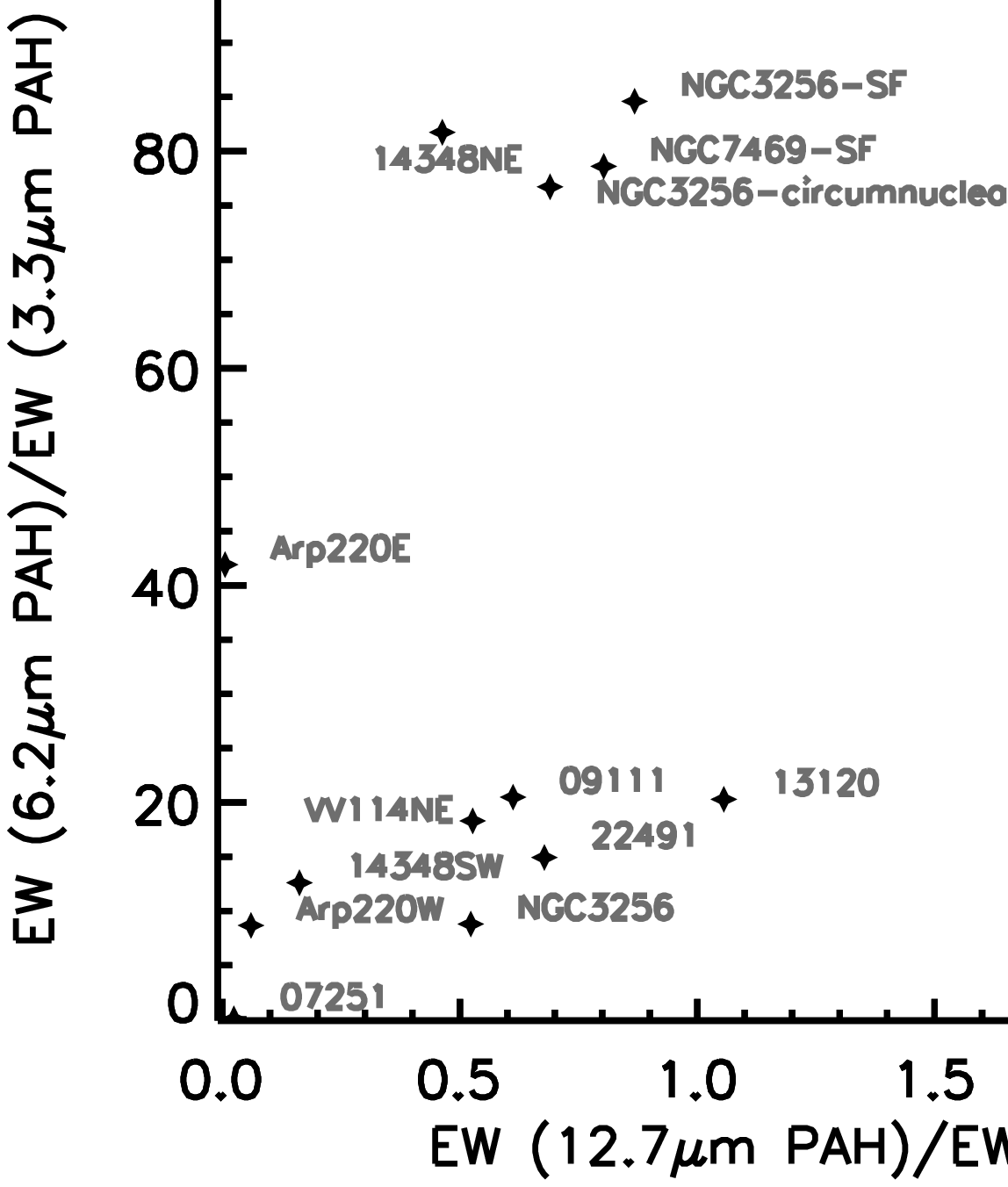}
\par}
\caption{ EW(6.2\,$\mu$m)/EW(3.3\,$\mu$m) vs EW(12.7\,$\mu$m)/EW(11.3\,$\mu$m) for the sample.} 
\label{check}
\end{figure}

\section{JWST colour-colour diagrams for selecting deeply obscured nuclei at various redshifts}
\label{additional_filters}
Here we investigate combination of different filters to those presented in Fig. \ref{filters} in Section \ref{filters_sect} which will enable us to select deeply obscured nuclei at various redshifts below z$=3$. In Fig. \ref{filters_variousz} we present various combination of filters to identify deeply obscured nuclei at redshift 1.5 and 2. 

Furthermore, at z$\sim$0.9 the ratios between the F560W and F1000W, F1130W and/or F1280W are effective selecting deeply obscured nuclei. Locally at z$\sim$0 buried nuclei could be selected using JWST/NIRCam filters (e.g. the ratio between F335M (or F356W) over F444W (or 480M)).

\begin{figure}[ht!]
\centering
\par{
\includegraphics[width=9.7cm]{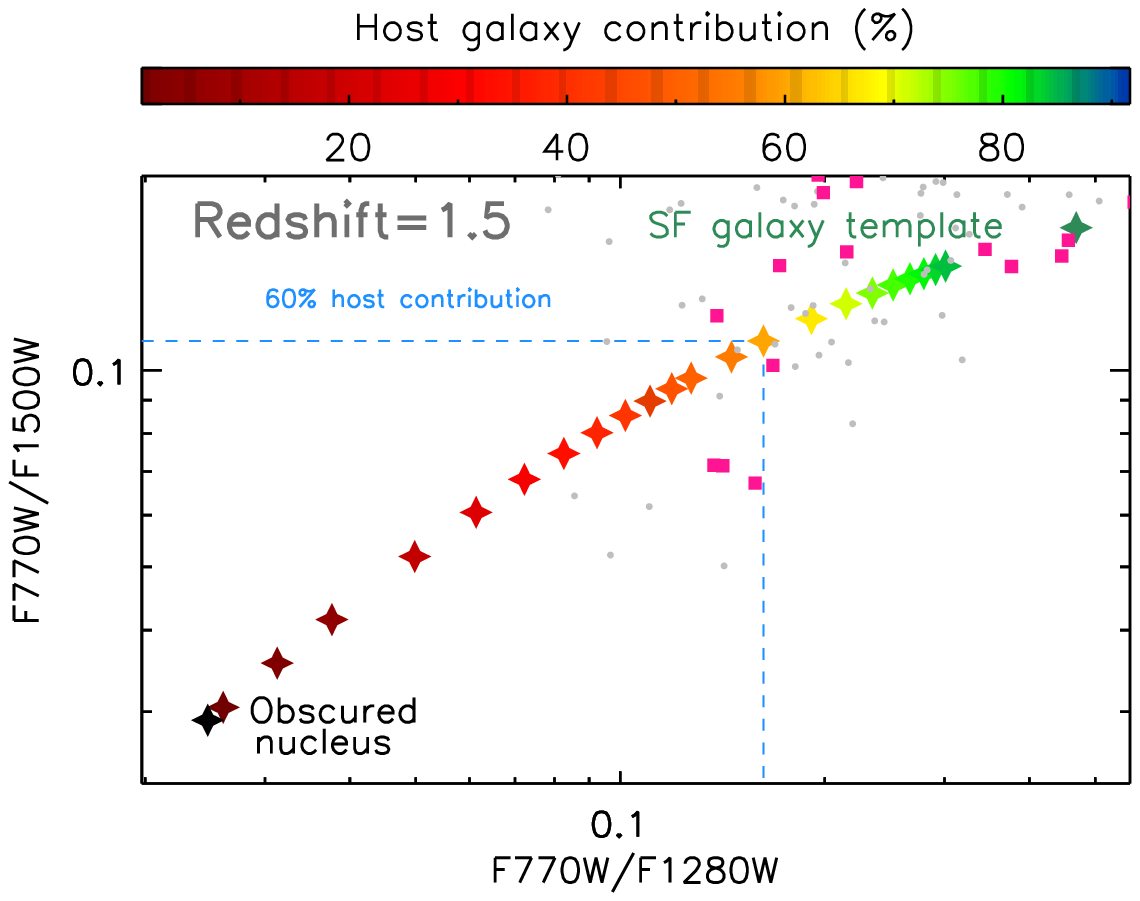}
\includegraphics[width=9.7cm]{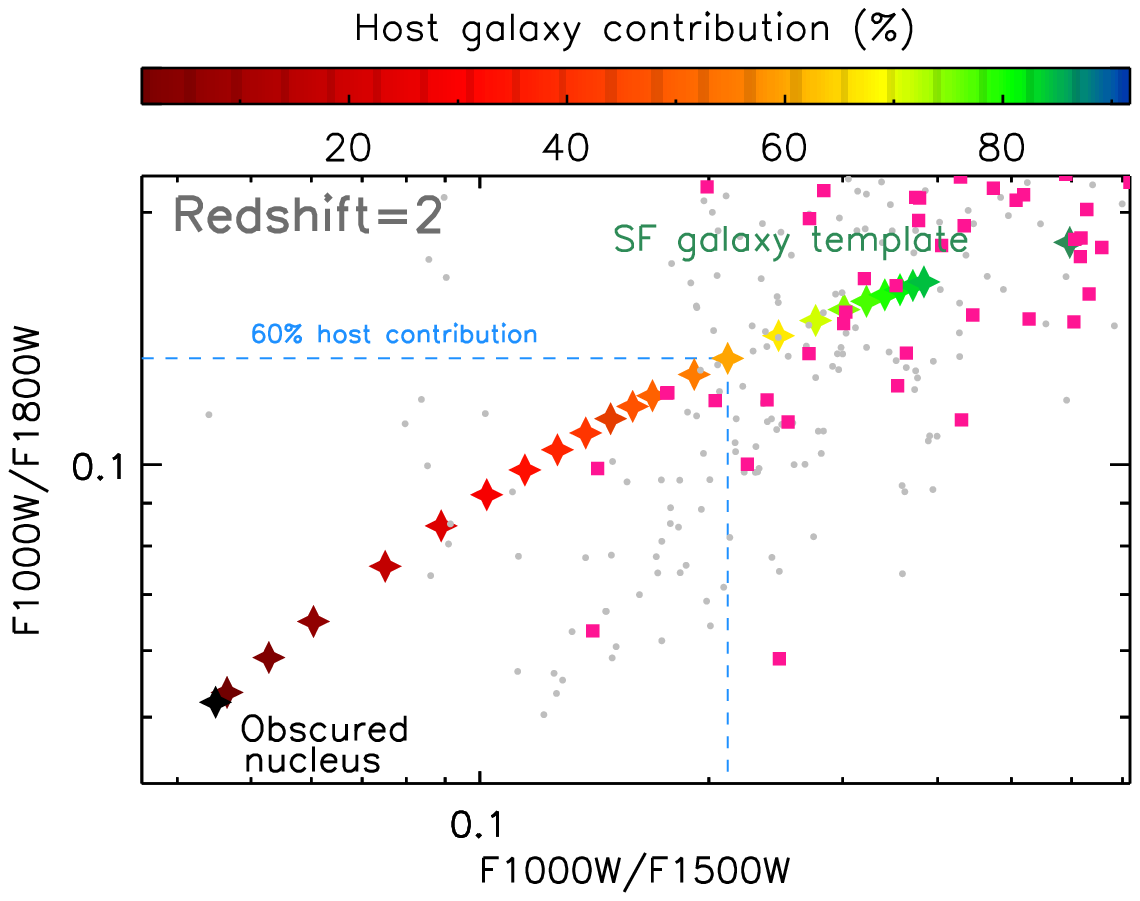}
\par}
\caption{ From top to bottom pannels: Colour-colour diagram (using the continuum) for selecting deeply obscured nuclei using JWST/MIRI filters z$=$1.5 and z$=$2. Colour-coded stars correspond to the host galaxy contribution of the simulated spectra shown in Fig. \ref{Synth_con}. The blue dashed lines denotes the region where sources have a contribution of the buried nuclei greater than 40\%. Grey circles correspond to CEERS sources with redshift measurements. Magenta squares represent those sources at redshift $\sim$1.25-1.75 (top panel) and 1.75-2.25 (bottom panel).} 
\label{filters_variousz}
\end{figure}

\end{appendix}

\end{document}